\documentclass[11pt]{article}
\usepackage{graphicx} 
\usepackage[a4paper,margin=1in]{geometry}
\usepackage{amsmath}
\usepackage{hyperref}
\usepackage{bigints}
\usepackage[utf8]{inputenc}
\usepackage{subcaption}
\usepackage{amsfonts,amssymb, amscd, amsmath,amsthm,amsfonts,epsfig,enumerate, yfonts}
\usepackage[font=small,skip=0pt]{caption}
\usepackage{float}
\usepackage{tikz}
\usetikzlibrary{positioning}
\usepackage[font=small,skip=7pt]{caption}
\usepackage{authblk}
\title{Gravitational Lensing and Image Distortion by Buchdahl Inspired Metric in $\mathcal{R}^2$ Gravity}
\author[1]{Shafia Maryam}
\author[1]{Mubasher Jamil\thanks{email:  mjamil@sns.nust.edu.pk}}
\affil[1]{\textit{School of Natural Sciences, National University of Sciences and Technology, H-12, 44000 Islamabad, Pakistan}}
\author[2]{Mustapha Azreg-A\"{\i}nou\thanks{email: azreg@baskent.edu.tr}}
\affil[2]{\textit{Ba\c{s}kent University, Engineering Faculty, Ba\u{g}l\i ca Campus, 06790-Ankara, Turkey}}
\author[3]{Zoe C S Chan\thanks{email:
bravychancs@gmail.com}}
\affil[3]{\textit{Department of Physics, Xiamen University Malaysia, 43900 Sepang, Malaysia}}

\date{September 2024}

\begin{document}

\maketitle
\begin{abstract}
We investigate gravitational lensing by \textit{special} Buchdahl inspired metric with the Buchdahl parameter $\tilde{k}$. In strong deflection limit, we derive the deflection angle analytically for the light rays that diverge as photons approach the photon sphere. These are then used in order to compute the angular image positions  modeling supermassive black holes, Sgr A* and M87* as  lenses. The Einstein rings for the outermost relativistic images are also depicted here alongside observational constraints on $\tilde{k}$ by the Einstein radius and lens mass. Constraints on $\tilde{k}$  are obtained modelling black holes ( Sgr A* and M87*) and Canarias Einstein ring.
In weak deflection limit, the analytic expression of deflection angle of the subject asymptotically flat metric in $\mathcal{R}^2$ gravity is determined using the Gauss Bonnet theorem. Considering M87* as a lens, weak deflection angle is used to study the image magnification and image distortion for primary and secondary images. It is shown that image distortion satisfies the hypothesis of Virbhadra. Moreover, it is seen that our general expression of deflection angle reduces, as a special case, to the deflection angle of Schwarzschild metric in both weak and strong deflection limits.
\end{abstract}

\section{Introduction}
Soon after the development of Theory  of General Relativity, the phenomenon of gravitational lensing was observed by Eddington  \cite{dyson1920ix}, confirming the predictions made by Einstein. Gravitational lensing had emerged from being a theoretical curiosity to a tool of modern astrophysics in research for over a century. 
The bending of light rays due to presence of gravity and winding of the photons around a black hole before the rays return to asymptotic region are well known since the works of Darwin \cite{darwin1959gravity}, Atkinson \cite{atkinson1965light} and Chankrasekhar \cite{chandrasekhar1983mathematical}.  Observation of distant astronomical objects by compact objects acting as gravitational lenses was made possible by the revolutionary work of Ellis and Virbhadra. They developed gravitational lens equation and numerically studied Schwarzschild lensing in strong limits \cite{virbhadra2000schwarzschild}. Later, Bozza extended their work and found analytical expressions of deflection angle for spherically symmetric black holes \cite{bozza2002gravitational}. The deflection angle of Kiselev black hole in the strong gravitational field was computed in \cite{younas2015strong}. They investigated  the positions and magnification of relativistic images in WDL (weak deflection limit). Then, Tsukamoto also worked on the analytical expression of strong gravitational lensing \cite{tsukamoto2017deflection} for spherically symmetric and static spacetime. With progressing works, an expression of strong gravitational lensing  with lensing coefficients $\Bar{a}$ and $\Bar{b}$ was established \cite{bozza2002gravitational} as follows

\begin{equation}\label{1}
\hat{\alpha}(b)=-\Bar{a}\log \left(\frac{b}{b_c}-1\right)+\Bar{b}+\mathcal{O}((b-b_c)\log(b-b_c)).
\end{equation}
Gibbons and Werner introduced the application of Gauss Bonnet theorem (GBT) on static, spherically symmetric spacetimes in order to find the defection angle of light in weak gravitational field \cite{gibbons2008applications}. Their work was confined for the asymptotic source and observer. They determined the analytical expression of deflection angle for Schwarzschild spacetime in weak field using their approach. Moreover, M. Bettinelli, M. Simioni and others reported a discovery of an optical Einstein ring which led to substantiation of the luminous source \cite{bettinelli2016canarias}. This instigated the interest of researchers in observing the Einstein ring. Two of the present authors studied the trajectory of a free particle around charged black holes \cite{azreg2020orbital}. They  studied variations in deflection angle by gravitational lensing because of the parameters involved in Einstein-{\AE}ther theory.  Recently,  magnification and angular position of images of  GUP-corrected-Schwarzschild spacetime (S-GUP) in the vicinity of weak gravitational lensing has been studied \cite{hoshimov2024weak}. The authors used the EHT  constraints of black holes M87* and Sgr A* to investigate lensing observables.

The Event Horizon Telescope (EHT) achieved a worthy milestone in the history of astrophysical studies by uncovering the first image of the black hole M87* \cite{event2019first} in the center of Messier 87 in the year 2019. Following this scientific feat, the first image of the black hole Sagittarius A* located at the heart of our Milky Way galaxy was unveiled on 12$^{th}$ May, 2022 \cite{akiyama2022first,event2023first,akiyama2022first3,akiyama2022first4}. The mass of black holes M87* and Sgr A* is inferred as $(6.5 \pm 0.7) \times10^9\textit{M}_\odot$ and $4.0^{+1.1}_{-0.6} \times10^6\textit{M}_\odot$, respectively.

Buchdahl instigated a search for static spherically symmetric metrics in vacuo using pure $\mathcal{R}^2$ action \cite{buchdahl1962gravitational}. Despite being very close to goal, unfortunately he abandoned to find analytical solution for his ordinary differential equation. Later, it remained unnoticed by researchers for over six decades until Nguyen advanced his work and obtained solutions to Buchdahl's ordinary differential equation. He recently obtained a class of static spherically symmetric solutions to field equations in pure $\mathcal{R}^2$ gravity, very fairly named ``Buchdahl-inspired metrics'', in \cite{nguyen2022beyond}. Continuing the successful work, an asymptotically flat metric was also proposed. The analytical expression for $special$ Buchdahl-inspired metric (SBI) was described for the first time in \cite{nguyen2023beyond}.  This is a very interesting metric involving a Buchdahl parameter $\tilde{k}$. It is detailed in work that for different values assigned to $\tilde{k}$, the metric supports different geometries.  For $\tilde{k}=0$, $\tilde{k}\in (-\infty,-1)\cup (0,+\infty)$ and $\tilde{k}\in (-1,0)$, it behaves like  Schwarzschild spacetime, a  naked singularity and a traversable Morris Thorne wormhole, each in order. For a very specific choice $\tilde{k}=-1$, SBI metric acts like some non-Schwarzschild structure \cite{nguyen2023traversable}. Furthermore, the observational tests of $\mathcal{R}^2$ spacetime in galactic center of Milky Way galaxy was carried out \cite{yan2024observational}.

In this paper, we will follow the approach of Tsukamoto \cite{tsukamoto2017deflection} to guide us in deriving the bending angle expressions of \textit{special} Buchdahl-inspired metric in strong gravitational field. Also, we will extend the foundational work of Gibbons and Werner \cite{gibbons2008applications} by applying their concept on this considered spacetime in order to find deflection angle in weak gravitational field. Throughout this paper, we adopt the geometric units $c=1=G$ in computations, unless mentioned otherwise. 

This paper is structured as follows: Section \ref{s2} contains an overview of \textit{special} Buchdahl-inspired metric. Section \ref{s3} showcase analytical expressions of this spacetime in strong limit alongside observables of relativistic images and constrains of the $\tilde{k}$ parameter against observational results of several Einstein rings. In section \ref{s4}, we determine the expression of weak deflection angle using Gauss Bonnet theorem. Magnification and image distortion of primary and secondary images will also be discussed.  We conclude the work with section \ref{s5}.

\section{\textit{Special} Buchdahl-inspired (SBI) metric: An overview}\label{s2}
The \textit{special} Buchdahl-inspired metric, characterised by a static, spherically symmetric vacuum configuration, was introduced  in \cite{nguyen2023beyond} as
\begin{equation}\label{2}
ds^2=\biggl|1-\frac{r_s}{r}\biggl|^{\tilde{k}}\biggl\{-\biggl(1-\frac{r_s}{r}\biggl)dt^2+\biggl(\frac{\rho(r)}{r}\biggl)^4\frac{dr^2}{1-\frac{r_s}{r}}+\biggl(\frac{\rho(r)}{r}\biggl)^2r^2d\Omega^2\biggl\},
\end{equation}
where the function $\rho(r)$ is described by
\begin{equation}\label{3}
\biggl(\frac{\rho(r)}{r}\biggl)^2=\frac{\zeta^2\big|1-\frac{r_s}{r}\bigl|^{\zeta-1}}{\bigl(1-s\bigl|1-\frac{r_s}{r}\bigl|^{\zeta}\bigl)^2}\biggl(\frac{r_s}{r}\biggl)^2,
\end{equation}
and $d\Omega^2=d\theta^2+\sin^2{\theta}\,d\phi^2$ is the conventional representation of a metric on the 2-sphere. The script $s=\pm1$ is set for the sign of $\left(1-\frac{r_s}{r}\right)$. The parameter $\tilde{k}$ is the new  Buchdahl parameter and $r_s$ plays the role of Schwarzscild radius\footnote{Note that for SBI metric, $r_s$ is scaled by a factor of $\frac{1}{1+\tilde{k}}$ such that $r_s=\frac{2GM}{(1+\tilde{k})c^2}$ \cite{zhu2024observational}.}
\begin{equation}\label{4}
\tilde{k}:=\frac{k}{r_s}, \hspace{1cm} \zeta:=\sqrt{1+3\tilde{k}^2}, \hspace{1cm}.
\end{equation}
both $\tilde{k}$ and the parameter $\zeta$ are dimensionless. Our prime focus is the computation of deflection angle and studying lensing observables, thus the region when $r>r_s$ is considered. Hence,  in the line element \eqref{2}  $s=+1$ shall be studied here. The metric is alternatively represented as
\begin{equation}\label{5}
ds^2=-\mathcal{A}(r)dt^2+\mathcal{B}(r)dr^2+\mathcal{C}(r)d\Omega^2,
\end{equation}
where 
\begin{equation}\label{6}
\begin{split}
	&f(r):=1-\frac{r_s}{r}\,,\\
	&\mathcal{A}(r)=f(r)^{\tilde{k}+1} ,
	\\&
	\mathcal{B}(r)=f(r)^{\tilde{k}-3}\frac{\zeta^4f(r)^{2\zeta}}{\big[1-f(r)^{\zeta}\big]^4}\biggl(\frac{r_s}{r}\biggl)^4,
	\\&
	\mathcal{C}(r)=f(r)^{\tilde{k}}\bigg(\frac{\zeta^2r_s^2f(r)^{\zeta-1}}{\big[1-f(r)^{\zeta}\big]^2}\bigg)= \mathcal{A}(r)\bigg(\frac{\zeta^2r_s^2f(r)^{\zeta-2}}{\big[1-f(r)^{\zeta}\big]^2}\bigg).
\end{split}
\end{equation}
The general spherically symmetric metric \eqref{5} is said to be asymptotically flat for all values of $\tilde{k}$ when 
\begin{equation}\label{7}
\begin{split}
	&\lim_{r\to\infty} \mathcal{A}(r)=1,
	\\&
	\lim_{r\to\infty} \mathcal{B}(r)=1,
	\\&
	\lim_{r\to\infty} \frac{\mathcal{C}(r)}{r^2}=1.
\end{split}
\end{equation}
The SBI spacetime delineated  in \eqref{6} reduces to Minkowski spacetime, for large distance $r\to\infty$, and mathematically satisfies the conditions given in \eqref{7}. Thus, the considered SBI metric is asymptotically flat.

\section{Strong gravitational lensing}\label{s3}
\begin{figure}[H]
\centering
\includegraphics[width=0.60\linewidth]{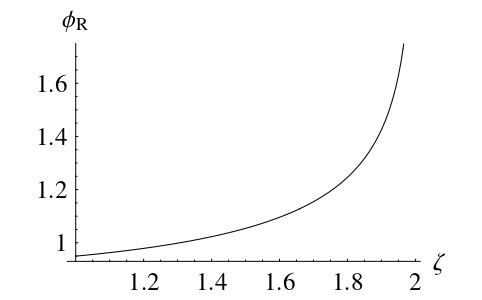}
\caption{\footnotesize{$\phi_R({\zeta})$~\eqref{30} versus $\zeta$ with $\phi_R({1})=2\ln [6(2-\sqrt{3})]$, the Schwarzschild value.}}
\label{Fig:reg}
\end{figure}

\begin{figure}
\centering
\begin{tikzpicture}
	\node (img1)  {\includegraphics[width=10cm]{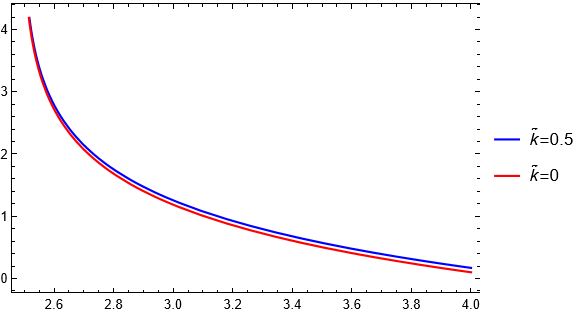}};
	\node[below=of img1, node distance=0cm, xshift=-0.8cm,yshift=0.7cm,font=\color{red}] {$b$};
	\node[left=of img1, node distance=0cm,,rotate=90, anchor=center,yshift=-0.7cm,font=\color{red}] {$\hat{\alpha}$};
\end{tikzpicture}
\caption{The strong deflection angle $\hat{\alpha}$ is plotted vs impact parameter $b$ for two values of Buchdahl parameter $\tilde{k}$. The red curve represents the case of SBH.}
\label{Fig:1}
\end{figure}

The spacetime under consideration is spherically symmetric which allows to set $\theta=\pi /2$ to find the equatorial deflection angles. The Lagrangian of a particle with 4-velocity $u^\mu$ is defined as 
\begin{equation}\label{8}
\mathcal{L}=\frac{1}{2} g_{\mu\nu} \dot{x}^\mu \dot{x}^\nu,
\end{equation}
here overdot represents differentiation with respect to an affine parameter $\lambda$. 
The Euler-Lagrange equation is described by 
\begin{equation}\label{9}
\frac{\partial \mathcal{L}}{\partial x^\mu}=\frac{d}{d\lambda}\frac{\partial \mathcal{L}}{\partial \dot{x}^\mu}
.\end{equation}
By using \eqref{9}, the equations of motion for a test particle (for $\mu=0$ and $\mu=3$)  in SBI spacetime \eqref{5} become
\begin{equation}\label{11}
\begin{split}
	&\dot{t}=\frac{\varepsilon}{\mathcal{A}(r)}\equiv \frac{\varepsilon}{f(r)^{\tilde{k}+1}},
	\\&
	\dot{\phi}=\frac{\ell}{\mathcal{C}(r)} \equiv \frac{\ell}{f(r)^{\tilde{k}}\left[\frac{\zeta^2r_s^2f(r)^{\zeta-1}}{\big[1-f(r)^{\zeta}\big]^2}\right]}.
\end{split}
\end{equation}
Here, $\varepsilon$ and $\ell$ represent the conserved energy of light rays and conserved angular momentum respectively. Now, using the null geodesic condition $ds^2=0$ and introducing the impact parameter $b:=\ell/\varepsilon$, we obtain 
\begin{equation}\label{12}
-\mathcal{A}(r)\dot{t}^2+\mathcal{B}(r)\dot{r}^2+\mathcal{C}(r)\dot{\phi}^2=0,
\end{equation}
\begin{equation}\label{13}
\begin{split}
	\dot{r}^2&=V_{\text{eff}}(r) \equiv \frac{1}{\mathcal{B}(r)}\left[\mathcal{A}(r)\dot{t}^2-\mathcal{C}(r)\dot{\phi}^2\right],
	\\&=\ell ^2\Bigl\{\frac{1}{b^2}\frac{1}{\mathcal{A}(r)\mathcal{B}(r)}-\frac{1}{\mathcal{B}(r)\mathcal{C}(r)}\Bigl\}.
\end{split}
\end{equation}
Using \eqref{6} in \eqref{13} gives the radial component of geodesic equation for SBI spacetime  as
\begin{equation}\label{14}
\dot{r}^2= \frac{\ell ^2 r^4f(r)^{2-2\zeta-2\tilde{k}} \big[1-f(r)^\zeta\big]^4}{\zeta^4r_s^4}\bigg[\frac{1}{b^2}-\frac{f(r)^{2-\zeta} \big[1-f(r)^\zeta\big]^2}{\zeta^2 r_s^2}\bigg]. 
\end{equation}
Consider an incoming light ray from a source at the asymptotic region. It reaches a minimum distance $r_{tp}$ from the black hole and then turn back in another direction towards an observer in the asymptotic region. The impact parameter can be found in terms of this closest distance $r_{tp}$ by imposing the condition of vanishing kinetic energy of light particles as follows 
\begin{equation}\label{15}
\frac{\dot{r}^2}{\ell^2}\biggl|_{r=r_{tp}}=0.
\end{equation}
Solving for $b$ brings the expression of the impact parameter at the turning point $r_{tp}$ to the form 
\begin{equation}\label{17}
b(r_{tp}) =\frac{\zeta r_s f(r_{tp})^{\frac{\zeta}{2}-1}}{1-f(r_{tp})^{\zeta}}=\sqrt{\frac{\mathcal{C}(r_{tp})}{\mathcal{A}(r_{tp})}}.
\end{equation}
When the closest approach distance $r_{tp}$ is located at the maximum of effective potential $V_{\text{eff}}$, this distance is called the radius of a photon sphere, denoted by ${r_c}$. Theoretically, ${r_c}$ gives the only bound orbit that photons could take and photons are captured by the black holes for $r<r_c$. This radius of photon sphere can be computed as given in \cite{atkinson1965light}, \cite{hsieh2021strong} and \cite{gan2021photon} using 
\begin{equation}\label{18}
\frac{d V_{\text{eff}}(r)}{dr}\biggl|_{r={r_c}}=0.
\end{equation}
We evaluate the derivative of \eqref{14} with respect to $r$ being cognizant of the finding that the impact parameter is a function of $r_{tp}$. Using the fact that $V_{\text{eff}}(r)\big|_{r={r_c}}=0$, we obtain
\begin{equation}\label{19}
\frac{d V_{\text{eff}}(r)}{dr}\bigg|_{r={r_c}}=\frac{\ell^2}{\mathcal{B}\mathcal{C}}\big[(\ln\mathcal{C})'-(\ln\mathcal{A})']\bigg|_{r={r_c}}=0.
\end{equation}
Using the last line in~\eqref{6}, $(\ln\mathcal{C})'=(\ln\mathcal{A})'+(\zeta -2)\big[\ln f(r)\big]'-2\big\{\ln\big[1-f(r)^\zeta\big]\big\}'$, yields $(\zeta -2)\big[\ln f(r)\big]'\big|_{r={r_c}} = 2\big\{\ln\big[1-f(r)^\zeta\big]\big\}'\big|_{r={r_c}}$, that is, 
\begin{equation}\label{rc}
\zeta -2+(\zeta +2)f(r_c)^\zeta=0.
\end{equation}
Knowing that $f(r_c)>0$ (because $r_c>r_s$), this last equation admits a solution only if\footnote{Constraints on $\tilde{k}$ have shown that $\zeta$ is closer to unity \cite{zhu2024observational}.} $\zeta <2$, yielding the radius of the photon sphere for SBI metric as
\begin{equation} \label{20}
{r_c}=\frac{r_s}{1-\left(\frac{2-\zeta}{2+\zeta}\right)^\frac{1}{\zeta}}=\frac{r_s}{1-p}\,,\qquad p:=\bigg(\frac{2-\zeta }{2+\zeta }\bigg)^{\frac{1}{\zeta }}.
\end{equation}
The corresponding critical impact parameter $b_c$ can be determined using  \eqref{17} for $r_{tp}={r_c}$, we obtain 
\begin{equation}\label{21}
b_c=\frac{r_s(\zeta+2)}{2}\Big(\frac{2+\zeta}{2-\zeta}\Big)^{\frac{2-\zeta}{2\zeta}}.
\end{equation}
As a function of closest approach $r_{tp}$, the deflection angle is defined as follows 
\begin{equation}\label{22}
\hat{\alpha}(r_{tp})=\phi(r_{tp})-\pi,
\end{equation}
where $\phi(r_{tp})$ can be found using \eqref{11} and \eqref{14}  as follows:
\begin{equation}\label{23}
\begin{split}
	\phi(r_{tp})&=2\int_{r_{tp}}^\infty \frac{dr}{\mathcal{C}}\sqrt{\frac{\mathcal{A}\mathcal{B}}{\frac{1}{b^2}-\frac{\mathcal{A}}{\mathcal{C}}}},
	\\&=2\int_{r_{tp}}^\infty \frac{\zeta r_s\,dr}{r^2\sqrt{f(r_{tp})^{2-\zeta} \big[1-f(r_{tp})^\zeta\big]^2-f(r)^{2-\zeta} \big[1-f(r)^\zeta\big]^2}}.
\end{split}
\end{equation}
It is obvious that the Taylor series about $r=r_{tp}$ of the expression under square root has no independent term. The coefficient of $(r-r_{tp})$ in the series is generally different from zero and it vanishes if $r_{tp}= r_c$. To demonstrate that it is better to introduce a new radial variable proportional to $r-r_{tp}$ at $r=r_{tp}$ as $z=r-r_{tp}$ or, preferably as done in \cite{tsukamoto2017deflection},
\begin{equation}\label{24}
z=1-\frac{r_{tp}}{r}.
\end{equation}
Then the integral $\phi(r_{tp})$ is expressed as
\begin{equation}\label{25}
\phi(r_{tp})=\frac{2\zeta r_s}{r_{tp}}\int_{0}^1 \frac{dz}{\sqrt{f(r_{tp})^{2-\zeta} \big[1-f(r_{tp})^\zeta\big]^2-f(r)^{2-\zeta} \big[1-f(r)^\zeta\big]^2}}=\int_0^1{h(z,r_{tp})}dz,
\end{equation}
where $r=r_{tp}/(1-z)$ and
\begin{equation}\label{26}
{h(z,r_{tp})}=\frac{2\zeta r_s}{r_{tp}}\frac{1}{\sqrt{f(r_{tp})^{2-\zeta} \big[1-f(r_{tp})^\zeta\big]^2-f(r)^{2-\zeta} \big[1-f(r)^\zeta\big]^2}}.
\end{equation}
We have
\begin{multline}\label{series}
f(r_{tp})^{2-\zeta} \big[1-f(r_{tp})^\zeta\big]^2-f(r)^{2-\zeta} \big[1-f(r)^\zeta\big]^2=\frac{r_s (r_{t p}-r_s)^{1-\zeta } [1-f(r_{t p})^{\zeta }] [\zeta -2+(\zeta +2) f(r_{t p})^{\zeta }]}{r_{t p}^{2-\zeta }}~ z\\
-\frac{r_s^2
	\big(\frac{r_{t p}}{r_{t p}-r_s}\big)^{\zeta } \{2 [1-f(r_{t p})^{\zeta }]^2-3 [1-f(r_{t p})^{2 \zeta }] \zeta +[1+f(r_{t p})^{2 \zeta }] \zeta ^2\}}{2 r_{t
		p}^2}~ z^2 +\cdots,
\end{multline}
and using~\eqref{rc} we see that the coefficient of $z$ vanishes if we replace $r_{tp}$ by $r_c$. This shows that the integral~\eqref{25} has a divergent term proportional to $\ln z$ in the limit $r_{tp}\to r_c$ of strong deflection ($b\to b_c$).
In this limit, we separate the regular part from the divergent part of the integral as follows \cite{tsukamoto2017deflection}
\begin{equation}\label{29}
\phi_R({r_c})=\phi_R({\zeta})=\int_0^1{[h(z,{r_c})-h_D(z,{r_c})]} dz\,,
\end{equation}
where, using~\eqref{6} and \eqref{series}, $h_D(z,r_{c})=2/z$. Now, we introduce the new variable $y$ as
\begin{equation}\label{y}
y := \bigg[1-\bigg(\frac{2-\zeta }{2+\zeta }\bigg)^{\frac{1}{\zeta }}\bigg]z+\bigg(\frac{2-\zeta }{2+\zeta }\bigg)^{\frac{1}{\zeta }}=(1-p)z+p,\qquad p:=\bigg(\frac{2-\zeta }{2+\zeta }\bigg)^{\frac{1}{\zeta }},
\end{equation}
in terms of which we write~\eqref{29} as
\begin{equation}\label{30}
\phi_R({r_c})=	\phi_R({\zeta})=2\zeta\int_{p}^1 \bigg\{\frac{1}{\sqrt{\frac{4 \zeta ^2 p^2}{4-\zeta ^2}-y^{2-\zeta } \left(1-y^{\zeta }\right)^2}}-\frac{1}{\zeta
	(y-p)}\bigg\} dy\,.
\end{equation}
This integral cannot be evaluated in terms of elementary functions for all values of $\zeta$. For $\zeta =1$ (corresponding to Schwarzschild metric with $\tilde{k}=0$), we obtain the exact value $\phi_R({1})=2\ln [6(2-\sqrt{3})]$. Numerical integration of~\eqref{30} yields the graph shown in Fig.~\ref{Fig:reg}.
The coefficient $\Bar{a}$ in~\eqref{1} has the general expression \cite{tsukamoto2017deflection}
\begin{equation}\label{35}
\Bar{a}=\sqrt{\frac{2 \mathcal{B}({r_c}) \mathcal{A}({r_c})}{\mathcal{C}''({r_c})\mathcal{A}({r_c})-\mathcal{C}({r_c}) \mathcal{A}''({r_c})}},
\end{equation}
where $\mathcal{A}(r)$, $\mathcal{B}(r)$ and $\mathcal{C}(r)$ are given in~\eqref{6}. This coefficient for SBI metric reduces to
\begin{equation}\label{36}
\Bar{a}=1.
\end{equation}
The next task is to find the coefficient $\Bar{b}$ which is described generally by \cite{tsukamoto2017deflection}
\begin{equation}\label{37}
\Bar{b}=\Bar{a}\ln{\Big[({r_c})^2\Big(\frac{\mathcal{C}''({r_c})}{\mathcal{C}({r_c})}-\frac{\mathcal{A}''({r_c})}{\mathcal{A}({r_c})}\Big)\Big]}+\phi_R({r_c})-\pi.
\end{equation}
On substituting $\mathcal{A}(r)$, $\mathcal{C}(r)$ and their double derivatives with respect to $r$ along with the value of $\Bar{a}=1$ and the radius of the photon sphere~\eqref{20}, we obtain
\begin{equation}\label{38}
\Bar{b}=\ln\Big[\frac{1}{2} (2+\zeta)(2-\zeta)^{1-\frac{2}{\zeta}} \big[(2-\zeta)^\frac{1}{\zeta}-(2+\zeta)^\frac{1}{\zeta}\big]^2\Big]+\phi_R({r_c})-\pi \,,
\end{equation}
where $\phi_R({r_c})$ is evaluated in \eqref{30}.
The deflection angle in the strong deflection limit of SBI metric is thus described by \eqref{1} with lensing coefficient determined in \eqref{36} and \eqref{38}. For Buchdahl parameter $\tilde{k}=0$, the expressions of  $\Bar{a}$  and $\Bar{b}$ reduce to
\begin{equation}\label{39}
\begin{split}
	&\Bar{a}=1.
	\\&
	\Bar{b}=\ln[216(7-4\sqrt{3})] -\pi\equiv-0.40023.
\end{split}
\end{equation}
These are the strong deflection limit coefficients for Schwarzschild spacetime~\cite{tsukamoto2017deflection}. Thus our calculations are consistent by yielding older results. The plot of deflection angle $\hat {\alpha}$ against impact parameter $b$ for two different behaviors of SBI metric is given in Fig \ref{Fig:1}.  The deflection angle decreases with the increase in impact parameter. 

\subsection{Relativistic images and observables}
In asymptotically flat SBI spacetime, the strong deflection observables can be studied using the lens equation as given in \cite{bozza2002gravitational}, 
\begin{equation}\label{40}
\beta=\theta - \frac{D_{LS}}{D_{OS}} \Delta \alpha_n,
\end{equation}
where $\beta$ and $\theta$ represents the  position of source and the position of image, respectively.  $D_{LS}$ denotes the distance between the black hole lens to be considered and the  light source.  $D_{OS}$ symbolizes the distance between the source and observer. Also, $\Delta \alpha_n$ is the offset of deflection angle $\hat{\alpha}$. When the photon orbits around the lens $n$ times then the offset of deflection angle is obtained upon  subtracting $2n\pi$ from the total deflection angle $\hat{\alpha}$: $\Delta \alpha_n=\hat{\alpha}-2n\pi$. Moreover, the impact parameter, in relation with the image position is given by $ b=D_{OL}\tan\theta \approx D_{OL} \theta$. Then, deflection angle \eqref{1} becomes 
\begin{equation}\label{41}
\hat{\alpha}(b)=-\Bar{a}\log \left(\frac{D_{OL} \theta}{b_c}-1\right)+\Bar{b}.
\end{equation}

If the light source, lens and observer are all aligned collinearly ($\beta=0$), this gives rise to an Einstein ring. The deflection angle together with \eqref{40} yields the \emph{zeroth} order angular position of the $nth$ image as
\begin{equation}\label{42}
\theta _n^0=\frac{b_c}{D_{OL}}\left(1+e_n\right),
\end{equation}
where
\begin{equation}\label{43}
e_n=\exp\left(\frac{\Bar{b}-2n\pi}{\Bar{a}}\right).
\end{equation}
So, the Einstein ring for the  $nth$ relativistic images is given by
\begin{equation}\label{44}
\theta _n^E=\frac{b_c}{D_{OL}}\left(1+e_n\right),
\end{equation}
Another important observable is the image magnification described as in\cite{bozza2002gravitational}  as
\begin{equation}
\mu_{n}=\frac{1}{\beta}\left[\frac{b_c(1+e_n)}{D_{OL}}\frac{D_{OS}b_c e_n}{D_{OL}D_{LS} \Bar{a}}\right].
\end{equation}
At sufficiently far distance $D_{OS}=2D_{LS}$, the magnification of n-loop images simplifies to
\begin{equation}\label{45}
\mu_{n}=\frac{1}{\beta}\left[{b_c(1+e_n)}\frac{2 b_c e_n}{D_{OL}^2 \Bar{a}}\right].
\end{equation}
Considering the case when the outermost image $\theta_1$ can be  separated and all the other relativistic images are gathered at asymptotic image position $\theta_\infty$, these two observable characteristics are described as \cite{bozza2002gravitational}
\begin{equation}\label{45a}
\theta_\infty=\frac{b_c}{D_{OL}},
\end{equation}
\begin{equation}\label{46}
\overline{s}=\theta_1-\theta_\infty\approx \theta_\infty \exp{\left(\frac{\Bar{b}-2n\pi}{\Bar{a}}\right)}.
\end{equation}

In order to study gravitational lensing  in strong deflection limit for SBI metric, two massive black holes are taken separately as a lens here. First, we model the supermassive black hole Sgr A* at the galactic center of Milky Way galaxy as a lens. The mass of Sgr A*  is approximated in recent studies \cite{islam2024strong} to be $4\times10^6 \textit{M}_\odot$ and the observer-lens distance is  0.008 Mpc. We numerically estimate the angular position of the outermost image $\theta_1$, the innermost image $\theta_\infty$ and the angular separation between these two images $\overline{s}$. To do so, \eqref{42}, \eqref{45a} and \eqref{46} are used with the deflection angle coefficients computed in \eqref{36} and \eqref{38}. 
We choose $\tilde{k}$ arbitrarily and calculate the corresponding Einstein radius. For $\tilde{k}=0$ (SBH), the Einstein radius for the outermost image is approximately 25.6783 $\mu as$. Similarly, the $\theta^{E}_1$ value for $\tilde{k}=0.03 $ (naked singularity) is 24.9271 $\mu as$ and for $\tilde{k}=-0.04 $ (Morris Thorne wormhole), it is 26.7419  $\mu as$. We tabulate the results for $\tilde{k} \in [-0.03,0.03]$ in Table \ref{tab:TabSA} for seven different values.

\begin{table}[htp]
\centering
\begin{tabular}{|c|c|c|c|c|c|c|c|}
\hline
$\tilde{k}$ & -0.03 & $-0.02$ & $-0.01$ & $0$ & $0.01$ & $0.02$ & $0.03$ \\
\hline
$b_c/r_s$ & $2.5628$ & $2.5826$ & $2.5942$ & $2.5981$ & $2.5942$ & $2.5826$ & $2.5628$ \\
\hline
$\Bar{a}$ & $1$ & $1$ & $1$ & $1$ & $1$ & $1$ & $1$ \\
\hline
$\Bar{b} $  & $-0.40002$ & $-0.40014$ & $-0.40021$ & $-0.40023$ & $-0.40021$ & $-0.40014$ & $-0.40002$ \\
\hline
$\theta_\infty$ $(\mu as)$ & $26.4359$ & $26.1681$ & $25.9049$ & $25.6462$ & $25.3919$ & $25.1419$ & $24.8959$ \\
\hline
$\theta_1$ $(\mu as)$ & $26.4690$ & $26.2008$ & $25.9373$ & $25.6783$ & $25.4237$ & $25.1733$ & $24.9271$ \\
\hline
$\overline{s}$ $(\mu as)$  & $0.03309$ & $0.03275$ & $0.03242$ & $0.03210$ & $0.03178$ & $0.03147$ & $0.03116$ \\
\hline
\end{tabular}
\caption{\footnotesize{Some relativistic images observables estimates for SBI spacetime in strong lensing using Sgr A* as a lens. Also lensing coefficients $\Bar{a}$ and $\Bar{b}$ for different values of Buchdahl parameter $\tilde{k}$ has been approximated.}}
\label{tab:TabSA}
\end{table}

\begin{figure}[htp]
\centering
\begin{tikzpicture}
	\node(imgA)
	{\includegraphics[scale=0.55]{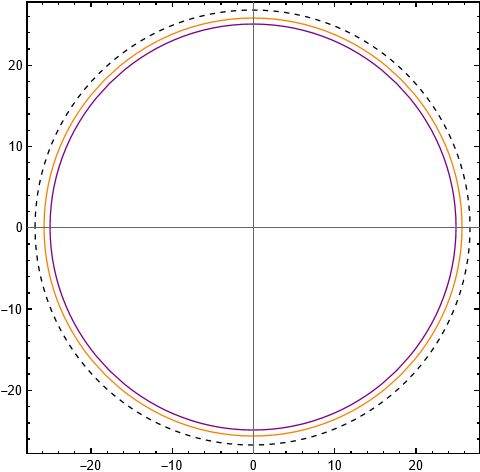}};
	\node[below=of imgA, node distance=0cm, yshift=1cm,font=\color{red}]{X} ;
	\node[left=of imgA, node distance=0cm, rotate=90, anchor=center,yshift=-0.7cm,font=\color{red}] {Y};
	\node[below=of imgA, node distance=0cm, yshift=0.4cm,font=\color{black}]{(a)};
	
	\node[right=of imgA, node distance=1cm, yshift=0cm](imgD)   {\includegraphics[scale=0.55]{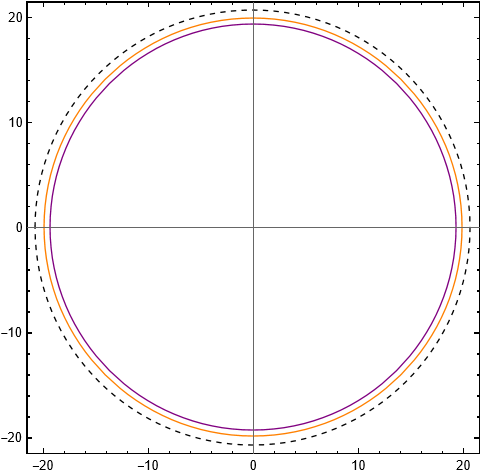}
	};
	\node[below=of imgD, node distance=0cm, yshift=1cm,font=\color{red}] {X};
	\node[left=of imgD, node distance=0cm, rotate=90, anchor=center,yshift=-0.7cm,font=\color{red}] {Y};
	\node[below=of imgD, node distance=0cm, yshift=0.4cm,font=\color{black}]{(b)};
	
\end{tikzpicture}
\caption{\footnotesize{Formation of the outermost relativistic images by SBI spacetime for $\tilde{k}=0$ (orange), $\tilde{k}=0.03$ (purple) and $\tilde{k}=-0.04$ (dashed)  is demonstrated modelling massive black holes as lens. Plot (a) and (b) represents Einstein ring  considering Sgr A* and M87*  as lens, respectively.}}
\label{fig:2}
\end{figure}
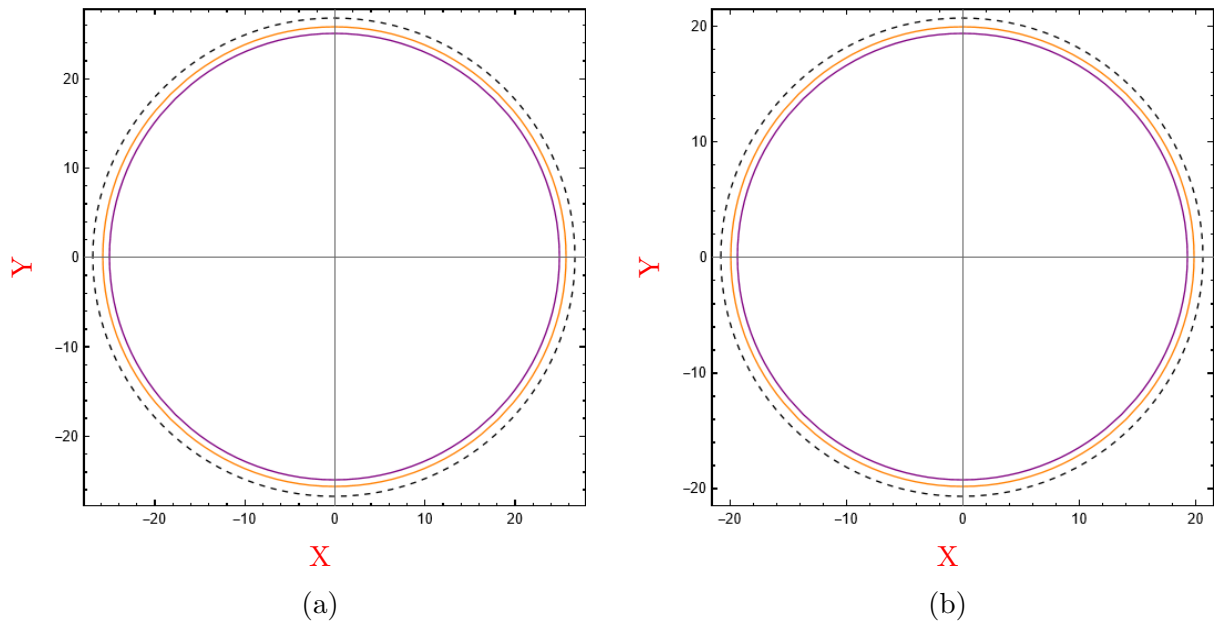

Next we model the M87 black hole of the Messier 87 galaxy as a lens for SBI spacetime and investigate its lensing observables. The mass of M87$^*$ is $6.5\times10^9 \textit{M}_\odot$ and the observer to lens distance is $16.8 Mpc$ \cite{islam2024strong}. We follow the same procedure as we did for Sgr A*. The numerical computations are tabulated in the Table \ref{tab:Tab M87}. Here, our Einstein ring radii for the outermost image are 19.8701 $\mu as$, 19.2888 $\mu as$ and 20.6932 $\mu as$ for $\tilde{k}=0, 0.03$ and $-0.04$ respectively. The Einstein rings for angular position of the outermost relativistic images of both Sgr A* and M87* are illustrated in the Fig. \ref{fig:2}.

\begin{table}[htp]
\centering
\begin{tabular}{|c|c|c|c|c|c|c|c|}
\hline
$\tilde{k}$ & -0.03 & $-0.02$ & $-0.01$ & $0$ & $0.01$ & $0.02$ & $0.03$ \\
\hline
$b_c/r_s$ & $2.5628$ & $2.5826$ & $2.5942$ & $2.5981$ & $2.5942$ & $2.5826$ & $2.5628$ \\
\hline
$\Bar{a}$ & $1$ & $1$ & $1$ & $1$ & $1$ & $1$ & $1$ \\
\hline
$\Bar{b} $  & $-0.40002$ & $-0.40014$ & $-0.40021$ & $-0.40023$ & $-0.40021$ & $-0.40014$ & $-0.40002$ \\
\hline
$\theta_\infty$ $(\mu as)$ & $20.4564$ & $20.2491$ & $20.0455$ & $19.8453$ & $19.6485$ & $19.4550$ & $19.2647$ \\
\hline
$\theta_1$ $(\mu as)$ & $20.4820$ & $20.2745$ & $20.0706$ & $19.8701$ & $19.6731$ & $219.4794$ & $19.2888$ \\
\hline
$\overline{s}$ $(\mu as)$  & $0.02560$ & $0.02534$ & $0.02509$ & $0.02484$ & $0.02459$ & $0.02435$ & $0.02411$ \\
\hline
\end{tabular}

\caption{\footnotesize{Approximates for observables and strong lensing coefficients of SBI metric  in strong deflection limit, considering M87* black hole as a lens, for different values of  $\tilde{k}$. }}
\label{tab:Tab M87}
\end{table}

Our results for Einstein ring with $\tilde{k}=0$ (SBH) closely resembled the Einstein ring in  \cite{islam2024strong}. From Tables \ref{tab:TabSA} and \ref{tab:Tab M87}, we notice that the Einstein rings for wormholes ($1 < \tilde{k} < 0$) are larger than its SBH counterpart whilst  Einstein rings for $k>0$, corresponding to a naked singularity are smaller. These comparisons roughly aligned with the results for spherically symmetric wormholes \cite{dey2008gravitational} and naked singularity \cite{babarNSLens}. Thus, our results are consistent with SBH. 

\begin{figure}[htp]
\centering
\begin{tikzpicture}
	\node(imgA)
	{\includegraphics[scale=0.45]{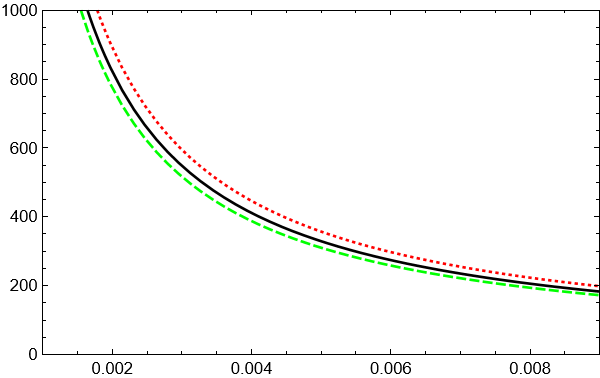}};
	\node[below=of imgA, node distance=0cm, yshift=1cm,font=\color{red}]{$\beta$ $(\mu as)$} ;
	\node[left=of imgA, node distance=0cm, rotate=90, anchor=center,yshift=-0.7cm,font=\color{red}] {$\mu$};
	\node[below=of imgA, node distance=0cm, yshift=0.4cm,font=\color{black}]{(a)};
	
	\node[right=of imgA, node distance=1cm, yshift=0cm](imgD)   {\includegraphics[scale=0.55]{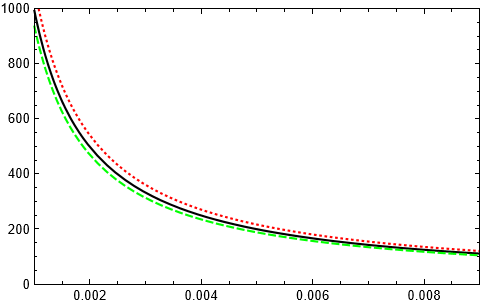}
	};
	\node[below=of imgD, node distance=0cm, yshift=1cm,font=\color{red}] {$\beta$ $(\mu as)$};
	\node[left=of imgD, node distance=0cm, rotate=90, anchor=center,yshift=-0.7cm,font=\color{red}] {$\mu$};
	\node[below=of imgD, node distance=0cm, yshift=0.4cm,font=\color{black}]{(b)};
	
\end{tikzpicture}
\caption{\footnotesize{Magnification of the outermost image by SBI spacetime for $\tilde{k}=0$ (black), $\tilde{k}=0.03$ (green) and $\tilde{k}=-0.04$ (red)  is demonstrated modelling massive black holes as lens. Plot (a) and (b) represents magnification  considering Sgr A* and M87*  as lens, respectively.}}
\label{fig:Mag}
\end{figure}
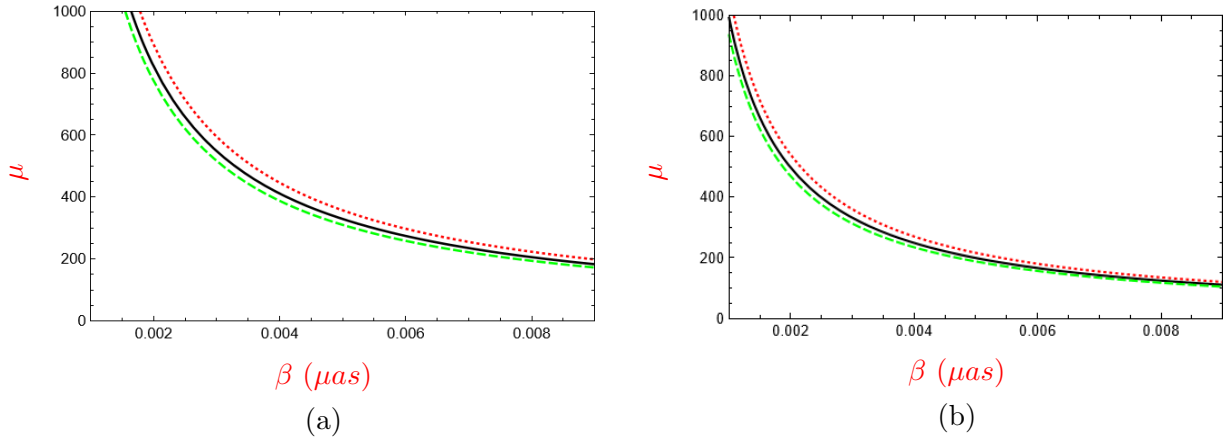

The magnification of the outermost image (assuming $n=1$) had been observed for three values of Buchdahl parameter. The graph plotted in Fig. \ref{fig:Mag} shows that magnification is maximum as the angular source position $\beta \to 0$ for both black holes. Thereby, the source must be perfectly aligned in order to maximize the magnification of the images.

\subsection{Mass enclosed within Einstein ring and constrains on \texorpdfstring{$\tilde{k}$}{TEXT} }
Einstein rings were also known for allowing measurement of mass that produced the lensing effect. The mass of a lens are estimated using observational data, typically with values of the Einstein radius and observer-lens distance. 
The practice of finding constraints for novel theoretical parameters (like $\tilde{k}$ in our case) provides framework for testing predictions of alternate-gravity models against observations. For the rest of this subsection, we will restore $r_s$ for SBI metric in its full form: $r_s=\frac{2GM}{(1+\tilde{k})c^2}$.

To start, we will constrain $\tilde{k}$ against observational values of the black hole mass and Einstein radius from EHT imaging for both Sgr A* and M87. Here, the Einstein radius is assumed to be equal to the outermost image value $\theta^{E}_1$. To get the formula for mass due to strong lensing, we substitute \eqref{21} into \eqref{44}. Setting $n=\Bar{a}=1$ and rearranging gives

\begin{equation}\label{eq:45}
M_1=\dfrac{c^2\theta_E D_{OL} (1+\tilde{k})}
        {G\,e^{\Bar{b}-2\pi}(2+\sqrt{1+3\tilde{k}^2})}
   \left(\dfrac{2+\sqrt{1+3\tilde{k}^2}}{2-\sqrt{1+3\tilde{k}^2}}\right)^{\frac{1}{2}-
   \frac{1}{\sqrt{1+3\tilde{k}^2}}},
\end{equation}
where $\theta_E$ is the Einstein radius. 

The measurement as given by EHT for Sgr A* black hole \cite{akiyama2022first} are $M=4.0^{+1.1}_{-0.6} \times10^6\textit{M}_\odot$ for the mass and $\theta_E = (25.90 \pm 1.15)$ $\mu as$ for the Einstein radius. Taking the measurement uncertainty into consideration, we can constrain $\tilde{k}$ to be 

\begin{equation}\label{C1}
-0.21 < \tilde{k} < 0.30.
\end{equation}

Similarly, the measurements for the M87 black hole \cite{event2019first} are $M=(6.5 \pm 0.7) \times10^9\textit{M}_\odot$ and $\theta_E = (21.0 \pm 1.5)$ $\mu as$. This leads to the constraint on $\tilde{k}$ to be 

\begin{equation}\label{C2}
-0.22 < \tilde{k} < 0.13.
\end{equation}

Next, we do the constraint test on some Einstein rings strongly lensed by early-type galaxies. We will not take into account the galaxies' ellipticity or details describing the incomplete ring image, which depends on the source location on the caustics in the source plane \cite{bettinelli2016canarias,narayan,8oClockArc}. As such, we simply model the galaxy as a circularly symmetric point mass lens producing a single complete ring image. Including the observer-source distance $D_{OS}$ and the lens-source distance $D_{LS}$, the Einstein radius under this model can be described by\footnote{For SBI metric, the change mass quantity $M$ is changed to $\frac{M}{(1+\tilde{k})}$.} 

\begin{equation}\label{m1}
\theta_E^2=\dfrac{4GM}{(1+\tilde{k})c^2}
           \dfrac{D_{LS}}{ D_{OL}D_{OS}},
\end{equation}
or simple rearrangement gives the modelled mass parameter as

\begin{equation}\label{m2}
M_2=\dfrac{\theta_E^2(1+\tilde{k})c^2}{4G}
           \dfrac{ D_{OL}D_{OS}}{D_{LS}}.
\end{equation}

The following constraints are approximated using both (\ref{m1}) and (\ref{m2}). The data given in the Canarias Einstein ring \cite{bettinelli2016canarias}: $\theta_E = (2.16'' \pm 0.13'')$, $D_{OL}= 951 h^{-1}$ Mpc, $D_{OS}= 1192 h^{-1}$ Mpc, $D_{LS}= 498 h^{-1}$ Mpc and $M=(1.86 \pm 0.23) \times10^{12}\textit{M}_\odot$. Note that $h$ is a numerical value often associated with cosmological models \cite{cosmoh} and is equal to $0.7$. The constraint produce by this data is 
\begin{equation}\label{C3}
 -0.22 < \tilde{k} < 0.27    .
\end{equation}
Another Einstein ring measured with the aforementioned model is dubbed the ``8 O' Clock Arc" \cite{8oClockArc}.
The given data are $\theta_E = (3.32'' \pm 0.16'')$, $D_{OL}= 752 h^{-1}$ Mpc, $D_{OS}= 1141 h^{-1}$ Mpc, $D_{LS}= 863 h^{-1}$ Mpc and $M=(1.92^{+0.19}_{-0.18}) \times10^{12}\textit{M}_\odot$. This constrains $\tilde{k}$ to be 
\begin{equation}\label{C4}
 -0.178 < \tilde{k} < 0.214   .
\end{equation}
Looking at \eqref{C1}, \eqref{C2}, \eqref{C3} and \eqref{C4}, they all have similar lower limits. The constrained values do not deviate far away from $\tilde{k}=0$ in agreement with \cite{zhu2024observational}.

\section{Weak gravitational lensing}\label{s4}
SBI metric represents spherically symmetric, asymptotically flat and static spacetime. In order to find deflection angle of the considered metric in weak deflection limits, we use Gauss Bonnet theorem (GBT) on optical metric adapting Gibbons and Werner's approach \cite{gibbons2008applications}.
\subsection{The optical metric}
Consider the SBI metric described by line element in \eqref{5}. Since it is spherically symmetric, we  set $\theta=\frac{\pi}{2}$ to restrict light rays to orbit in the equatorial plane. This reduces $d\Omega^2=d\theta^2+\sin^2\theta d\phi^2$ to $d\phi^2$. Now making  use of the null geodesic condition $ds^2=0$, this leads to the form of optical metric $dt^2=\overline{g}_{ij}dx^idx^j$ as 
\begin{equation}\label{e1}
dt^2=\overline{g}_{rr} dr^2+\overline{g}_{\phi\phi} d \phi^2,
\end{equation}
where the optical metric components read as 
\begin{equation}\label{e2}
\overline{g}_{rr}=\left(\frac{\zeta^2\big(1-\frac{r_s}{r}\bigl)^{\zeta-2}}{\bigl(1-\bigl(1-\frac{r_s}{r}\bigl)^{\zeta}\bigl)^2}\biggl(\frac{r_s}{r}\biggl)^2\right)^2 ,\hspace{1.75cm} \overline{g}_{\phi\phi}=\frac{\zeta^2\big(1-\frac{r_s}{r}\bigl)^{\zeta-1}}{\bigl(1-\bigl(1-\frac{r_s}{r}\bigl)^{\zeta}\bigl)^2}\frac{{r_s}^2}{1-\frac{r_s}{r}},
\end{equation}
emphasizing that the components of optical metric are $\overline{g}_{rr} \neq {g}_{rr}$ and $\overline{g}_{\phi\phi} \neq {g}_{\phi\phi}$.

\subsection{GBT by Gibbon and Werner's approach}
\begin{figure}[htb]
\centering
\begin{tikzpicture}
	\node (img1)  {\includegraphics[width=10cm]{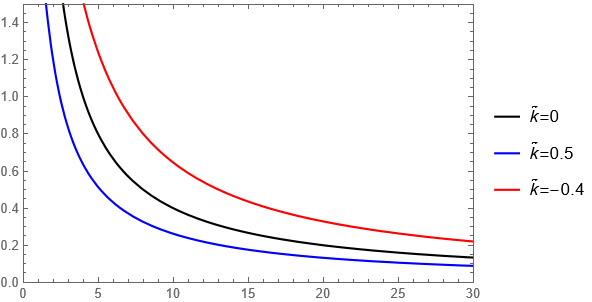}};
	\node[below=of img1, node distance=0cm, xshift=-0.8cm,yshift=0.7cm,font=\color{red}] {$b$};
	\node[left=of img1, node distance=0cm,,rotate=90, anchor=center,yshift=-0.7cm,font=\color{red}] {$\hat{\alpha}$};
\end{tikzpicture}
\caption{The weak deflection angle $\hat{\alpha}$ is plotted vs impact parameter $b$ for  different values of Buchdahl parameter $\tilde{k}$, corresponding to three different cases. The black curve shows the bending angle for SBH.}
\label{Fig:3}
\end{figure}

Gibbons and Werner demonstrated that weak deflection angle for asymptotically flat spacetime can be calculated in the weak field approximation as \cite{gibbons2008applications}
\begin{equation}\label{e3}
\hat{\alpha}=-\int\int_{D_\infty}\mathcal{K}dS,
\end{equation}
here the Gaussian curvature $\mathcal{K}$ is integrated over the domain $D_\infty$ of a surface $S$ where the observer and source are situated at  spatial infinity. Surface element is represented by $dS$. The Gaussian curvature provides information about curvature of the spacetime and its mathematical form can be found in \cite{kreyszig2013differential} (see page 147) as
\begin{equation}\label{e4}
\mathcal{K}=-\frac{1}{\sqrt{\Bar{g}_{rr}\Bar{g}_{\phi\phi}}}\left[\frac{\partial}{\partial r}\left(\frac{1}{\sqrt{\Bar{g}_{rr}}}\frac{\partial}{\partial r}(\sqrt{\Bar{g}_{\phi\phi}})\right)+\frac{\partial}{\partial \phi}\left(\frac{1}{\sqrt{\Bar{g}_{\phi\phi}}}\frac{\partial}{\partial \phi}(\sqrt{\Bar{g}_{rr}})\right)\right] .
\end{equation}
Since SBI metric  is independent of $\phi$, thereby Gaussian curvature reads
\begin{equation}\label{e5}
\mathcal{K}=-\frac{1}{\sqrt{\Bar{g}_{rr}\Bar{g}_{\phi\phi}}}\left[\frac{\partial}{\partial r}\left(\frac{1}{\sqrt{\Bar{g}_{rr}}}\frac{\partial}{\partial r}(\sqrt{\Bar{g}_{\phi\phi}})\right)\right].
\end{equation}
Now using the components of optical metric \eqref{e2} $\mathcal{K}$ becomes
\begin{equation}\label{e6}
\mathcal{K}=-\frac{\left(1-\frac{r_s}{r}\right)^2 \left[1-\left(1-\frac{r_s}{r}\right)^\zeta\right]^3}{r_s^2 \zeta^3 \left(1-\frac{r_s}{r}\right)^{\zeta/2}}.
\end{equation}
The surface element of  the optical metric can be desribed as
\begin{equation}\label{e7}
dS=\sqrt{\Bar{g}_{rr}\Bar{g}_{\phi \phi}}dr d\phi \equiv \frac{r_s^3 \zeta^3\left(1-\frac{r_s}{r}\right)^{3\zeta/2}}{r^2 \left(1-\frac{r_s}{r}\right)^3 \left[1-\left(1-\frac{r_s}{r}\right)^\zeta\right]^3} dr d\phi.
\end{equation}
Now for weak deflection angle make use of the straight line approximation which gives $r=\frac{b}{\sin  \phi}$~\cite{gibbons2008applications,javed2022weak,ovgun2019weak} and $0\leq \phi \leq \pi$. Also substitution of the optical Gaussian curvature from  \eqref{e6} and the surface element \eqref{e7} into  \eqref{e3} yields
\begin{equation}\label{e8}
\hat{\alpha}=\bigintss_0^\pi \bigintss_{\frac{b}{\sin \phi}}^\infty  \frac{r_s \left(1-\frac{r_s}{r}\right)^{\zeta}}{r^2 \left(1-\frac{r_s}{r}\right) } dr d\phi.
\end{equation}
Evaluation of this surface integral \eqref{e8} in \textit{Mathematica} gives 
\begin{equation}\label{e9}
\hat{\alpha}=\frac{\pi}{\zeta}-\frac{\pi}{\zeta} {_2F_1}\left[\frac{1-\zeta}{2}, -\frac{\zeta}{2}, 1, \frac{r_s^2}{b^2}\right] +\frac{2r_s}{b} {_PF_Q}\left[\biggl\{1,\frac{1-\zeta}{2}, 1-\frac{\zeta}{2}\biggl\},\biggl\{\frac{3}{2}, \frac{3}{2}\biggl\}, \frac{r_s^2}{b^2}\right].
\end{equation}
Here, ${_2F_1}$ is the hypergeometric function and ${_PF_Q}$ denotes the generalized hypergeometric function.  The deflection angle \eqref{e9} is plotted versus the impact parameter $b$ in Fig. \ref{Fig:3}. With the decrease in impact parameter, the deflection angle increases for the considered values of $\tilde{k}$. A Taylor series of~\eqref{e8} yields
\begin{equation}\label{e10b}
	\hat{\alpha}=2 \frac{r_s}{b}-\frac{\pi }{4}~(\zeta -1) \Big(\frac{r_s}{b}\Big)^2 +\frac{2}{9}~(\zeta -1) (\zeta-2) \Big(\frac{r_s}{b}\Big)^3 +(\zeta-1)~\mathcal{O}\left(\Big(\frac{r_s}{b}\Big)^4\right) \,,
\end{equation} 
where $r_s=2M(1+\tilde{k})$~\cite{azreg2023stationary}. For Buchdahl parameter $\tilde{k}=0$ ($\zeta =1$), the deflection angle in weak deflection limit \eqref{e9} transforms into 
\begin{equation}\label{e10}
\hat{\alpha}=\frac{2r_s}{b} \equiv \frac{4M}{b}.
\end{equation}
This is the weak deflection angle for Schwarzschild spacetime \cite{gibbons2008applications}. Thus, our new finding \eqref{e9} is consistent with SBH. 

\subsection{Image magnification and distortion}
To evaluate gravitational lensing phenomenon due to deflection of light in weak deflection limit, we employ the lens equation derived in~\cite{virbhadra2000schwarzschild} 
\begin{equation}\label{e11}
\tan \beta=\tan \theta - \alpha,
\end{equation}
where $\alpha$ is defined by
\begin{equation}\label{e12}
\alpha=\frac{D_{LS}}{D_{OS}} \tan{(\hat{\alpha}- \theta)}.
\end{equation}
Here,  $\hat{\alpha}$ is the deflection/bending angle. As before, the symbols $\beta$ and $\theta$ represents the angular position of the luminous source and position of the image, respectively. Also, $D_{LS}$  represents the lens-source distance and $D_{OS}$ signifies observer-source distance. 
The impact parameter can be rewritten in relation with the position of image as 
\begin{equation}\label{e13}
b=D_{OL} \sin \theta,
\end{equation}
where $D_{OL}$ denotes the observer-lens distance. 

In weak field approximations, the angular position of source is small and the lens equation reduces to \cite{virbhadra2022distortions} 
\begin{equation}\label{LE}
    \beta=\theta-\hat{\alpha}\mathcal{D},
\end{equation}
where $\hat{\alpha}$ is the weak deflection angle and parameter $\mathcal{D}$ is defined by $\frac{D_{LS}}{D_{OS}}$.
The total magnification $\mu$ and the components of radial $\mu_r$ and tangential $\mu_t$ magnification  are described as \cite{virbhadra2022distortions}  
\begin{equation}\label{e14}
\begin{split}
	&\mu=\mu_t \mu_r,
	\\&
	\mu_r=\left(\frac{d\beta}{d\theta}\right)^{-1},
	\\&
	\mu_t=\left(\frac{\sin\beta}{\sin\theta}\right)^{-1}.
\end{split}
\end{equation}

To verify if there is no image from the particular source missed in observations, Virbhadra \cite{virbhadra2022distortions} came up with a  hypothesis that there exists a distortion parameter such that the distortions of all images must be summed to obtain \textit{zero}. The distortion parameter of images is defined using tangential and radial magnifications \cite{virbhadra2022distortions} as follows
\begin{equation}\label{e15}
\Delta=\frac{\mu_t}{\mu_r},
\end{equation}
thus the logarithmic distortion parameter can be expressed as
\begin{equation}\label{e16}
\delta=\log_{10} |\Delta|.
\end{equation}
Statement of Virbhadra's hypothesis can mathematically be delineated as \cite{virbhadra2022distortions}
\begin{equation}\label{e17}
\sum _{i=1}^n \Delta_i=0,
\end{equation}
here $n$ quantifies the total number of images of light source formed by a single lens.

In this work, we model the massive black hole M87*  as a SBI lens. For  gravitational lensing by SBI spacetime in weak deflection limit, magnification of both primary and secondary images are investigated here.

\begin{figure}[htp]
\centering
\begin{tikzpicture}
	\node(imgA)
	{\includegraphics[scale=0.555]{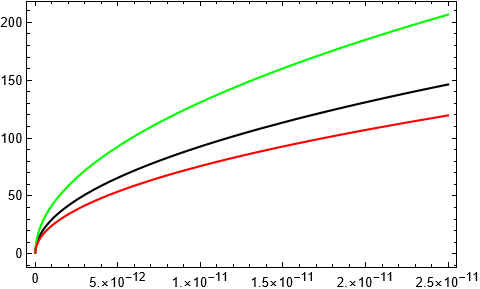}};
	\node[below=of imgA, node distance=0cm, yshift=1cm,font=\color{red}]{$M/D_{OL}$} ;
	\node[left=of imgA, node distance=0cm, rotate=90, anchor=center,yshift=-0.7cm,font=\color{red}] {$\mu _{pt}$};
	\node[below=of imgA, node distance=0cm, yshift=0.4cm,font=\color{black}]{(a)};
	
	\node[below=of imgA, node distance=1cm, yshift=-0.3cm](imgB)  {\includegraphics[scale=0.55]{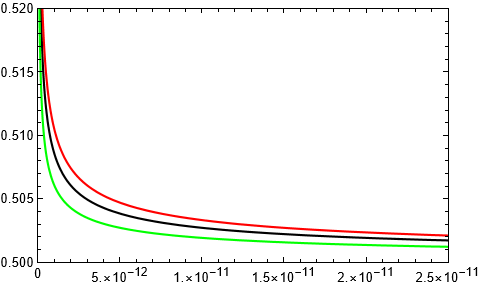}
	};
	\node[below=of imgB, node distance=0cm, yshift=1cm,font=\color{red}] {$M/D_{OL}$};
	\node[left=of imgB, node distance=0cm, rotate=90, anchor=center,yshift=-0.7cm,font=\color{red}] {$\mu_{pr}$}; 
	\node[below=of imgB, node distance=0cm, yshift=0.4cm,font=\color{black}]{(c)};
	\node[below=of imgB, node distance=1cm, yshift=-0.3cm](imgC)  {\includegraphics[scale=0.555]{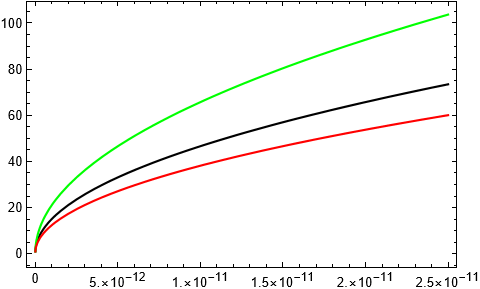}
	};
	\node[below=of imgC, node distance=0cm, yshift=1cm,font=\color{red}] {$M/D_{OL}$};
	\node[left=of imgC, node distance=0cm, rotate=90, anchor=center,yshift=-0.7cm,font=\color{red}] {$\mu_p$};
	\node[below=of imgC, node distance=0cm, yshift=0.4cm,font=\color{black}]{(e)};
	\node[right=of imgA, node distance=1cm, yshift=0cm](imgD)   {\includegraphics[scale=0.55]{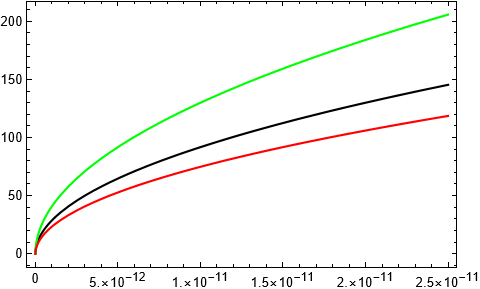}
	};
	\node[below=of imgD, node distance=0cm, yshift=1cm,font=\color{red}] {$M/D_{OL}$};
	\node[left=of imgD, node distance=0cm, rotate=90, anchor=center,yshift=-0.7cm,font=\color{red}] {\small{$|\mu_{st}|$}};
	\node[below=of imgD, node distance=0cm, yshift=0.4cm,font=\color{black}]{(b)};
	
	\node[right=of imgB, node distance=0cm, yshift=0cm](imgE)   {\includegraphics[scale=0.55]{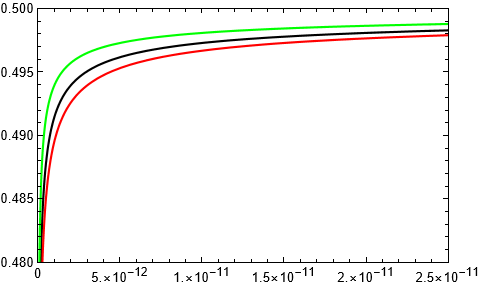}
	};
	\node[below=of imgE, node distance=0cm, yshift=1cm,font=\color{red}] {$M/D_{OL}$};
	\node[left=of imgE, node distance=0cm, rotate=90, anchor=center,yshift=-0.7cm,font=\color{red}] {$|\mu_{sr}|$};
	\node[below=of imgE, node distance=0cm, yshift=0.4cm,font=\color{black}]{(d)};
	\node[right=of imgC, node distance=1cm, yshift=0cm](imgF)
{\includegraphics[scale=0.55]{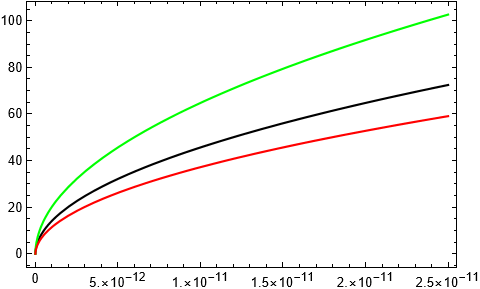}
	};
	\node[below=of imgF, node distance=0cm, yshift=1cm,font=\color{red}] {$M/D_{OL}$};
	\node[left=of imgF, node distance=0cm, rotate=90, anchor=center,yshift=-0.7cm,font=\color{red}] {$|\mu_s|$};
	\node[below=of imgF, node distance=0cm, yshift=0.4cm,font=\color{black}]{(f)};
\end{tikzpicture}
\caption{Magnifications are plotted for SBI spacetime in WDL. For primary images: the tangential magnification  $\mu_{pt}$, radial magnification  $\mu_{pr}$ and total magnification $\mu_p$ are graphed respectively in (a), (c) and (e) - against ratio  $M/D_{OL}$. For secondary images: the absolute tangential magnification $|\mu_{st}|$, the absolute radial magnification $|\mu_{sr}|$ and the absolute total magnification $|\mu_{s}|$ is plotted against $M/D_{OL}$ in (b), (d) and (f), each in order.
	Plot convention: The black, green and  red curves are respectively representing the plots for parameter values $\tilde{k}=0$, $-0.5$ and $0.5$.}
\label{Fig:4}
\end{figure}
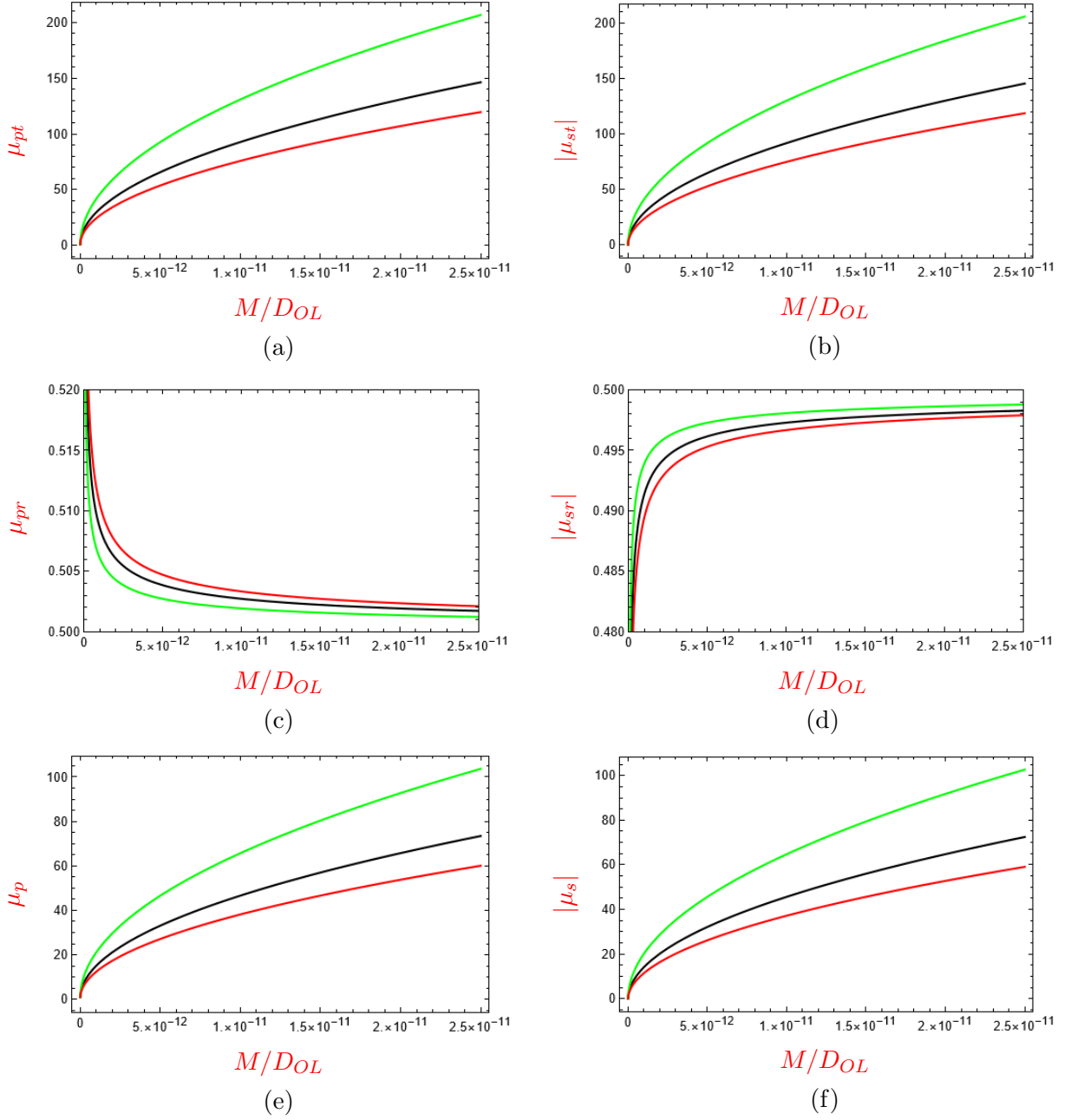

Solving the lens equation \eqref{LE} for $\theta$ yields the primary and secondary image positions as $\theta_p$ and $\theta_s$, respectively. Using these image positions the radial, tangential and total magnifications are obtained for both primary images and secondary images. 
The magnifications of primary and the secondary images are hence plotted in the Fig. \ref{Fig:4}. Plots in Fig. \ref{Fig:4}(a) and Fig. \ref{Fig:4}(b), respectively, show that primary tangential and absolute secondary tangential magnification increase when plotted against $M/D_{OL}$ for three different values of Buchdahl parameter. The plot in  Fig. \ref{Fig:4}(c) shows that the primary radial magnification at first drops rapidly and then at a slower rate when plotted against dimensionless parameter ratio $M/D_{OL}$. Furthermore, Fig. \ref{Fig:4}(d) depicts that the absolute secondary radial magnification rises promptly and then its rate of increase slows down as the parameter on the horizontal-axis increases. 

Then, the total primary magnification $\mu_{p}$ and the absolute total secondary magnification $|\mu_s|$ is plotted against $M/D_{OL}$ in Fig. \ref{Fig:4}(e) and Fig. \ref{Fig:4}(f), respectively. In these graphs it can be seen that both the total primary and absolute total secondary magnification increases. Here, if we compare the radial, tangential and total magnification curves for $\tilde{k}=0$ in all six plots with the respective plots of Schwarzschild lensing as in \cite{virbhadra2022distortions}, we can clearly see the similar behaviours of all the three magnifications. Henceforth, our plots are consistent with SBH magnification.

\begin{figure}[htp]
\centering
\begin{tikzpicture}
	\node(imgA)
	{\includegraphics[scale=0.56]{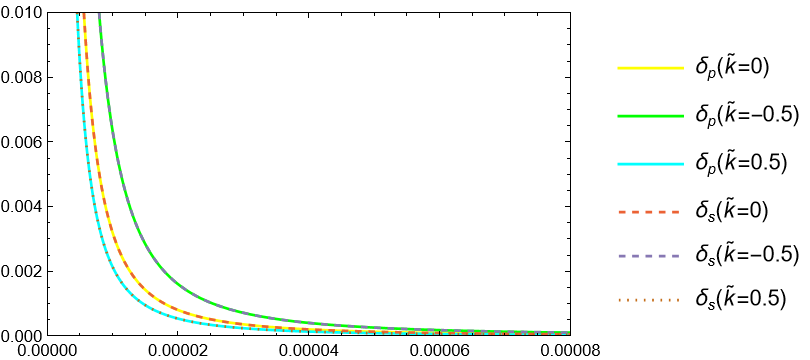}};
	\node[below=of imgA, node distance=0cm, yshift=1cm,xshift=-1.4cm, font=\color{red}]{$\beta$} ;
	\node[left=of imgA, node distance=0cm, rotate=90, anchor=center,yshift=-0.7cm, font=\color{red}] {$\delta$} ;
	\node[below=of imgA, node distance=0cm, yshift=0.4cm,xshift=-1.4cm,font=\color{black}]{(a)};
	
	\node[below=of imgA, node distance=1cm, yshift=-0.3cm](imgB)  {\includegraphics[scale=0.53]{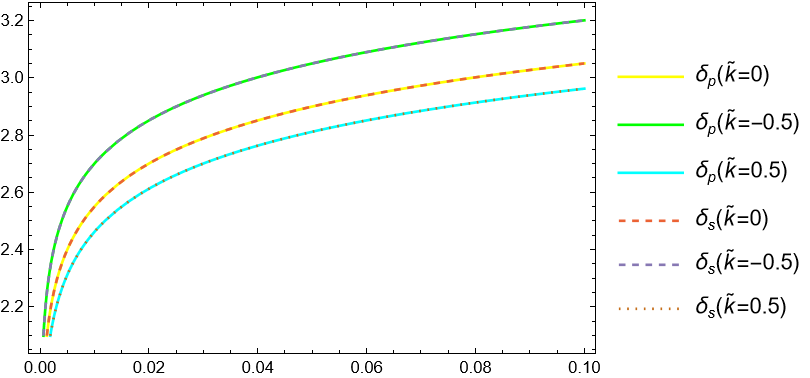}
	};
	\node[below=of imgB, node distance=0cm, yshift=1cm,xshift=-1.4cm,font=\color{red}] {$\mathcal{D}$};
	\node[left=of imgB, node distance=0cm, rotate=90, anchor=center,yshift=-0.7cm,font=\color{red}] {$\delta$}; 
	\node[below=of imgB, node distance=0cm, yshift=0.4cm, xshift=-1.4cm,font=\color{black}]{(b)};
	\node[below=of imgB, node distance=1cm, yshift=-0.3cm](imgC)  {\includegraphics[scale=0.555]{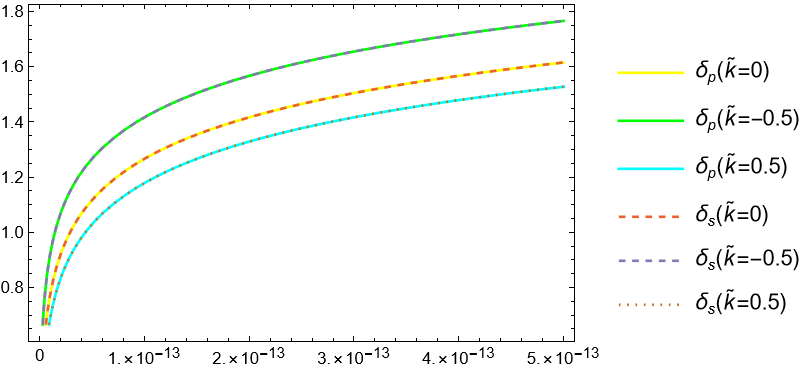}
	};
	\node[below=of imgC, node distance=0cm, yshift=1cm,xshift=-1.4cm,font=\color{red}] {$M/D_{OL}$};
	\node[left=of imgC, node distance=0cm, rotate=90, anchor=center,yshift=-0.7cm,font=\color{red}] {$\delta$};
	\node[below=of imgC, node distance=0cm, yshift=0.4cm,xshift=-1.4cm,font=\color{black}]{(c)};
\end{tikzpicture}
\caption{Logarithmic distortion is plotted for SBI spacetime in WDL. The logarithmic distortion of the primary images ($\delta_p$) and the secondary  images ($\delta_s$) is plotted against angular position of source $\beta$, distance parameter $\mathcal{D}$ and $M/D_{OL}$ in (a), (b) and (c), respectively.  
}
\label{Fig:5}
\end{figure}

In order to investigate the image distortion, we again model the M87* black hole of massive galaxy M87 as a SBI lens. The distortion parameter for SBI spacetime can be obtained in \textit{Mathematica} using the general expressions \eqref{e14} and \eqref{e16}. Next the logarithmic distortion parameter $\delta$ is plotted against some parameters in Fig.~\ref{Fig:5}. Both the primary logarithmic distortion $\delta_p$ and the absolute secondary logarithmic distortion $\delta_s$ are plotted against the other parameters which are the source position $\beta$ in Fig.~\ref{Fig:5}(a),
the dimensionless parameter $\mathcal{D}$ in Fig.~\ref{Fig:5}(b) and the ratio $M/D_{OL}$ in Fig.~\ref{Fig:5}(c).

The logarithmic distortion parameter for both primary and secondary images increases along with the increase in both $\mathcal{D}$ and $M/D_{OL}$. However, the parameter $\delta$ decreases in counter to the increase in $\beta$ for primary images and secondary images. For a given value of Buchdahl parameter $\tilde{k}$, the curves in all the plots of Fig.~\ref{Fig:5} are identical for the corresponding primary and secondary images. Thus, the total distortion parameter is zero, satisfying Virbhadra's hypothesis.

\section{Conclusion}\label{s5}
In this work, we determined the expressions of deflection angles due to weak and strong gravitational lensing
for asymptotically flat SBI spacetime.  The graphs
of the deflection angle in strong field are depicted for three cases supported by SBI metric i.e, Morris Thorne wormhole for $\tilde{k}\in (-1,0)$, SBH for $\tilde{k}=0$,  and a naked singularity for $\tilde{k} \in (-\infty,-1) \cup (0,\infty)$.  Additionally, we modeled the black holes Sgr A* and M87* as SBI lens to estimate the
positions of images and distance between the innermost and  the outermost image. Einstein rings are also plotted choosing three values from each case of SBI spacetime. Magnification of the outermost relativistic image has been investigated to maximize when angular source position $\beta \to 0$. The constraints on parameter $\tilde{k}$ modelling Sgr A* and M87* respectively suggest that $\tilde{k}\in (-0.21,0.30)$ and $\tilde{k} \in (-0.22,0.13)$. Also using the data of ``Canarias Einstein ring" and an ``8 O' Clock Arc", Buchdahl parameter is limited respectively as $\tilde{k} \in (-0.22,0.27)$ and $\tilde{k} \in (-0.178,0.214)$. Thus, the parameter $\tilde{k}$ is suitably constrain to not deviate far from zero.

Bending angle according to Gauss Bonnet theorem was calculated employing Gibbons and Werner’s approach in weak deflection limit. Tangential and total magnifications for primary and secondary images by model black hole M87* with mass $6.5 \times 10^9 \textit{M}_\odot$ are observed increasing when plotted against the ratio $M/D_{OL}$. While the radial magnification increases for
primary and decreases for secondary images. Moreover, the sum of distortion parameter of primary and secondary image vanishes which satisfies the hypothesis of Virbhadra \cite{virbhadra2022distortions}. Moreover, the results for analytical expression of deflection angle in strong and weak gravitational field are consistent with SBH for Buchdahl parameter $\tilde{k}=0$.

\bibliographystyle{IEEEtran}
\bibliography{References}
\end{document}